\documentclass[futureinternet,article,accept,pdftex,moreauthors]{Definitions/mdpi} 

\usepackage{listings}
\usepackage{url}
\usepackage{algorithm}
\usepackage{algorithmic}
\usepackage{graphicx}
\usepackage{mdframed}

\firstpage{1} 
\pubvolume{1}
\issuenum{1}
\articlenumber{0}
\pubyear{2024}
\copyrightyear{2024}
\datereceived{ } 
\daterevised{ } % Comment out if no revised date
\dateaccepted{ } 
\datepublished{ } 
\hreflink{https://doi.org/} % If needed use \linebreak
\Title{Explainable identification of similarities between entities for discovery in large text}

\TitleCitation{Text similarities}
\Author{Akhil Joshi $^{1}$, Sai Teja Erukude $^{1}$ and Lior Shamir $^{1,*}$\orcidA{}}
\AuthorNames{ Akhil Joshi, Sai Teja Erukude and Lior Shamir}
\AuthorCitation{Joshi, A.; Erukude, S.T.; Shamir, L.}

\address{%
$^{1}$ \quad Kansas State University, Manhattan, KS 66502} % ;  erukude.saiteja@gmail.com, akhil.joshi08@gmail.com, lshamir@mtu.edu } % \\
\corres{Correspondence: lshamir@mtu.edu}

\abstract{With the availability of virtually infinite number text documents in digital format, automatic comparison of textual data is essential for extracting meaningful insights that are difficult to identify manually. Many existing tools, including AI and large language models, struggle to provide precise and explainable insights into textual similarities. In many cases they determine the similarity between documents as reflected by the text, rather than the similarities between the subjects being discussed in these documents. This study addresses these limitations by developing an n-gram analysis framework designed to compare documents automatically and uncover explainable similarities. A scoring formula is applied to assigns each of the n-grams with a weight, where the weight is higher when the n-grams are more frequent in both documents, but is penalized when the n-grams are more frequent in the English language. Visualization tools like word clouds enhance the representation of these patterns, providing clearer insights. The findings demonstrate that this framework effectively uncovers  similarities between text documents, offering explainable insights that are often difficult to identify manually. % By focusing on document-to-document comparison, the method allows to explore complex textual data with greater precision. 
This non-parametric approach provides a deterministic solution for identifying similarities across various fields, including biographies, scientific literature,  historical texts, and more. Code for the method is publicly available.}

\begin{document}

\section{Introduction} 
\label{introduction}

In the modern digital landscape, the need for effective methods to analyze, compare and interpret large textual data is paramount. The ability to detect patterns of similarities between text documents can provide immediate insights that would otherwise be difficult to make. For instance, users who study certain political figures can use large amounts of text to identify certain similarities between them. Researchers in the medical domain can identify similarities or common causes between two clinical conditions by comparing a large corpus of medical records that describe these conditions. Similarly, health records associated with certain drugs or vaccines can be compared to identify symptoms or side effects, which can be analyzed even by scanning user-generated documents through social media. In the humanities domain, comparisons can be used to identify specific similarities between novels, literary styles, or characters. But despite the potential usefulness of the ability to quickly perform an explainable identification of similarities between documents, such algorithms have yet attracted relatively little attention. 

Automatic analysis of text has been studied intensively for tasks such as automatic document classification \citep{kowsari2019text,gasparetto2022survey,minaee2021deep,yu2008evaluation}, automatic translation of documents \citep{bar1960present,ablanedo2007efficacy}, document summarization \citep{campr2015comparing,carenini2013multi,ma2022multi}, and automatic generation of documents, which is often done through large language models \citep{wu2024large}. 

Text document comparison is a common feature in modern word processors, but its goal is merely to compare and match words and sentences in the body of the text. It cannot be used effectively for comparing specific topics of similarities discussed in the documents. Therefore, while automatic document comparison is commonly used feature, it is irrelevant to the task of identifying specific similarities between two or more entities as they are studied through the text documents.

One of the most common forms of identifying similarities between documents is automatic plagiarism detection \citep{parker1989computer,foltynek2019academic,naik2015review,chandere2021online,kulkarni2021analysis}. Other research efforts focused on similarities between scientific documents, including similarities in keywords and references \citep{ahlgren2009document}. Term extraction was also used to determine document similarities \citep{milios2003automatic}. Domains in which automatic document similarities are commonly used for include law \citep{bhattacharya2022legal,chen2022comparative} and accounting \citep{bochkay2023textual}. Using machine learning, the task of clustering was applied to text documents to identify the most similar documents inside a text corpus \citep{bezdan2021hybrid}. Similarities between text documents can also be done using stylistic elements of the writing \citep{shamir2021udat,swisher2023data}. 

Document similarity comparison can include string-based methods, knowledge-based methods, corpus-based method, and hybrid methods that combine different approaches \citep{prasetya2018performance}. String-based similarities aim at lower level comparison of the text by comparing two or more sequences of letters. For instance, that can be done by the minimum number of primitive operations required to transform one strong into the other \citep{wang2011fast,levenshtein1965binary,damerau1964technique}. The approach of edit distance has been used intensively also outside the task of text comparison, for instance in field of genetics to compare and identify matching DNA sequences \citep{mcginnis2004blast}. 

Knowledge-based methods use semantic networks to extract information and make conclusions from text. It uses pre-defined ontologies to make connections between meaning and words, often in specific predefined domain. The difference between two pieces of text can be measured through the path on the semantic network that needs to be taken in order to lead to a match \citep{meng2013review,rada1989development}. Another form is through collections of common terms or concepts \citep{huedo2006assessing,tversky1977features}.

Corpus-based similarity of text is based on applying similarities between sub-parts of the text, normally sentences, and a defined collection of text documents. The comparison to single yet very large set of text documents allows to find similarities to common sentences that appear in the corpus, which allows to identify similarities or to transform text documents \citep{landauer1997solution,gabrilovich2007computing,kuipers2013empowerment}.  

Automatic text similarity has been used intensively for the purpose of document clustering and for data mining in text \citep{alqahtani2021survey,xinwu2012new,ghaemi2009survey}. For instance, cosine similarity can be used by comparing a feature vector made of the similarities between different components \citep{basha2017improved}. Association rule is another common method to identify similarities between text by using certain association procedures \citep{kulkarni2016knowledge,manimaran2013survey,oliinyk2016stochastic}. 

While these methods have been useful for their designated purposes, they are less applicable for the purpose of discovering patterns of similarities between entities described in text documents. The similarity scores are used effectively for the purpose of text data mining, text information retrieval, and document clustering. However, they do not necessarily provide direct insights on the specific elements that assist to study similarities between entities. Also, these methods are often based on complex rules, and therefore not always explainable in the sense that the user can understand why different patterns express similarities between the documents. Existing AI and large language models can also be used for the task of identifying similarities between different entities, but they often provide generalized summaries that do not necessarily capture subtle similarities nuances.

Therefore, many of these current solutions struggle with delivering explainable insights that are trivial for the user to understand. This study seeks to enhance existing Natural Language Processing (NLP) tools by developing an explainable framework that can identify specific similarities between large text documents. Ultimately, these similarities should be communicated to the user in a clear and intuitive manner. The analysis is domain-independent, and can be applied without training a machine learning system, collecting data, or crafting semantic networks based on existing knowledge.

% This study emphasizes a deeper exploration of linguistic relationships to overcome these challenges.

By focusing on detailed comparisons, this framework aims to facilitate more informative assessments of similarities between pieces of text. The analysis is explainable, and allows the user to understand the meaning of the outcomes and reconstruct the path that led to the results. The proposed approach can therefore work in concert with other approaches such as solutions based on large language models, and provide additional insights that large language models often struggle to identify. Since the algorithm is deterministic and clear, users can easily evaluate and understand the nature of the output.

% Central to the framework is the integration of advanced NLP techniques, such as Named Entity Recognition (NER) and part-of-speech (POS) tagging, leveraging tools like SpaCy and NLTK. The study introduces a custom algorithm that computes weighted differences in n-gram frequencies, enhancing the detection of significant linguistic features.

%Organized into three main sections, the research details the methodologies for n-gram extraction, the framework's development, and the results, employing visualizations to illustrate key findings. This work contributes to the field of computational linguistics, providing a scalable approach for analyzing textual similarities and paving the way for further advancements in NLP applications.

\section{Automatic similarity analysis method}
\label{method}

The purpose of the method is to provide a quick overview of topics that are common between the two documents, and might therefore be relevant to similarities between the entities being studied. A simple comparison of words is not necessarily effective since it will provide a majority of terms that are common to the documents being compared, but provide no specific information about similarities between the entities. For that purpose, n-grams are used in combination with several other NLP tools.

\subsection{N-grams}
\label{ngram&nlp}

\begin{figure}[ht]
    \centering
    \includegraphics[scale=0.5]{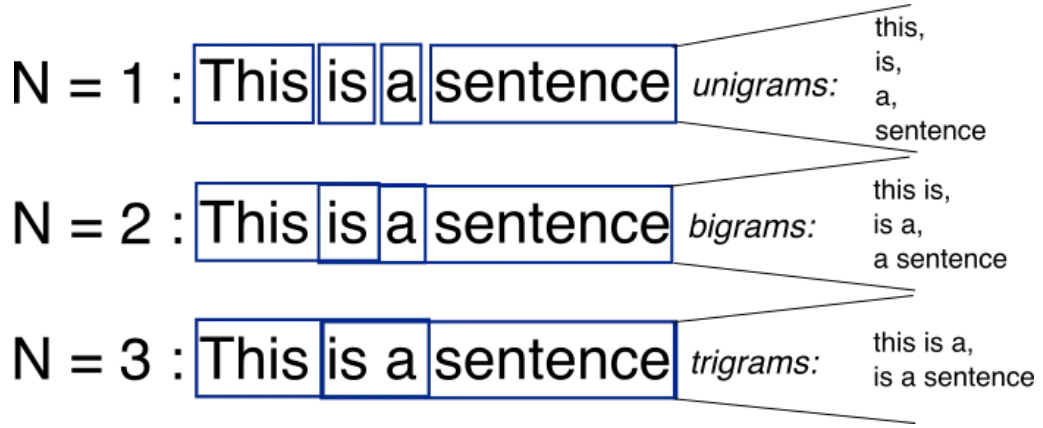}
    \caption{Examples of 1-grams, 2-grams, and 3-grams.}
    \label{fig_mean_lat_variation}
\end{figure}

In the context of NLP, n-grams are contiguous sequences of ’n’ items, typically words \citep{sidorov2014syntactic,robertson1998applications,damashek1995gauging,silva2024large}. These items can also be characters or syllables, depending on the granularity of the analysis. The value of ’n’ determines the order of the n-gram. In this study, the n-grams are key to understanding the similarities between texts by focusing on patterns in word sequences. They can be used to pinpoint common phrases and structures, which allows for a more informative comparison between documents. Figure~\ref{fig_mean_lat_variation} shows an example of 1-grams, 2-grams, and 3-grams.

Analysis through n-grams requires a step of pre-processing of the text data such as tokenization and lemmatization of the text. That was done with the two commonly used NLP libraries: {\it SpaCy} and {\it NLTK}. {\it SpaCy} is known for high efficiency and accuracy, making it optimal for real-time applications or tasks that involve the analysis of very large quantities of text \citep{schmitt2019replicable,altinok2021mastering,fantechi2021spacy}. One of the advantages of SpaCy in the context of this study is that it supports models of named entity recognition (NER) and dependency parsing. On the other hand, SpaCy’s limited focus on high-level processing requires additional tools for detailed corpus-based analysis.

{\it NLTK} is comprehensive library supporting diverse text processing techniques such as stemming, stopword removal, and part of speech tagging. While NLTK may be slower on large datasets, it offers capabilities for detailed analysis and preprocessing. In this study it is used for part of speech tagging and entity identification.

% Together, SpaCy and NLTK offer a complementary approach that addresses a broad range of NLP requirements by leveraging the strengths of each library. Rather than relying on a single tool, we combined the efficiency and pre-trained capabilities of SpaCy with the flexibility and research-oriented functionalities of NLTK, after evaluating other libraries that ultimately didn’t match the needs of this project. This dual-library setup capitalizes on the best of both worlds SpaCy’s speed and accuracy for tasks like Named Entity Recognition and dependency parsing, and NLTK’s customizable modules for n-grams and detailed text preprocessing. This combined approach provides a comprehensive toolkit, effectively balancing performance and depth to meet the project's objectives.

\subsection{Weighted n-gram analysis for identification of similarities}
\label{algorithm}

%This section presents the function \texttt{analyze\_common\_ngrams}, 

The analysis of identifiable similarities between documents of text is based on which identifying and assigning weights to n-grams that appear in both documents.  % The focus here is on the final iteration of the function, which integrates techniques like Part-of-Speech (POS) tagging and Named Entity Recognition (NER) to enhance the analysis \citep{jurafsky2019speech} \citep{manning1999foundations} \citep{cavnar1994ngram}. 
The analysis receives two chunks of text data as its input, and identifies n-grams that are common between them. For that purpose, the words needs to be converted to their stem form to identify similar n-grams. The is done through the {\it SpaCy} library, which provides mature automatic lemmatization of text.

The concept of n-grams is needed since many patterns of similarities can not be detected through a single word. For instance, in a comparison of two preeminent scientists, a possible pattern of similarity between them is that both wan a Nobel Prize. Since ``Nobel Prize'' is two words, the word ``Nobel'' and ``Prize'' will not be associated if each is used as a single word. Another example could be the term ``climate change'', where each of the two words has a meaning in its own, but when combined they reflect a different term. The use of n-grams can ensure that terms that are reflected by multiple words are also analyzed, and can be detected.

The primary weakness of this technique is that the vast majority of matching n-grams in two documents are irrelevant to the similarities between the subjects discussed in these documents. That is, while the text might be similar due to common words that appear in both of them, the similarity of the text does not provide any useful insights to the user. For instance, simple words or phrases such as ``it is'' might appear numerous times in the documents, but the fact that they appear in both documents does not provide any useful insights. In that sense the similarity between text documents provides similarity of the text, but not the similarity between the subjects that are being compared. To address that weakness, the analysis must be able to trim n-grams of common words that are frequent in both documents. Such n-grams are not expected to include useful insights, and will overwhelm the user with unimportant data, making it difficult to identify the true informative insights. That can be done by penalizing topics or words that are frequent in the English language. The following algorithm provides the high-resolution description of the method. 

%The \texttt{analyze\_common\_ngrams} function analyzes and assigns a weighted score to n-grams that appear in both documents being compared, operating as follows:

% Algorithm Pseudocode
\begin{algorithm}[H]
\caption{High-level algorithm for common n-grams analysis}
\begin{algorithmic}[1]
    \STATE \textbf{Input:} doc1\_data, doc2\_data, eng\_dict
    \STATE \textbf{Output:} weighted\_ngrams
    \STATE Analyze n-grams in \texttt{doc1\_data} and \texttt{doc2\_data}
    \STATE Identify common n-grams between documents
    \FOR{each common n-gram}
        \STATE Calculate weighted score based on frequency using \texttt{eng\_dict}
    \ENDFOR
    \STATE Apply SpaCy for noun chunk and named entity recognition
    \STATE Combine weighted scores of common n-grams from both documents
    \RETURN weighted\_ngrams
\end{algorithmic}
\end{algorithm}

% Arguments
\textbf{Arguments}
\begin{itemize}
    \item \texttt{doc1\_data} and \texttt{doc2\_data}: N-gram data, noun chunks, and named entity information for each document.
    \item \texttt{eng\_dict}: A dictionary of English word frequencies for significance calculation.
\end{itemize}

% Throughout our project, we undertook multiple iterations, each building on the last to refine our approach to n-gram analysis. We started with basic implementations, gradually integrating more sophisticated techniques. Each iteration contributed valuable insights, forming a solid foundation for our final solution. \\

The backbone of the method is the weighting of the n-grams as they are evaluate for their ability to provide useful similarities between two text documents. Such weighting can be conceptualized as the identification of n-grams that are common to the two documents being compared, but are not so common in the general language used in these documents. That is, the weight of an n-gram increases when the n-gram is common in the two document, but decreases when the n-gram is common in the language of the document. That provides an explainable method that can identify specific n-grams that indicate on similarities between documents.

In its most simple form, the weights can be determined using a basic frequency comparison:

\begin{equation}
    Weight = (f_1 - f_e)^2 \times (f_2 - f_e)^2,
\end{equation}
where \( f_1 \) and \( f_2 \) represent n-gram frequencies in the two documents, and \( f_e \) represents baseline frequency for common words derived from a large and representative Corpus of Contemporary American English \citep{davies2010corpus}. The weight can therefore be conceptualized as the frequency of the word in the document in which it appears compared to its frequency in the English language. Words that are more frequent in the document are assigned with a higher weight, but that weight is also reduced when the word is more frequent in the English language in general.

To allow adjusting the noise level through the weighting, a logarithmic scaling was applied as shown by Equation~\ref{logarithmic_scaling}. The logarithmic scaling provided a smoother weighting that further lowered the weight of n-grams that are common in the English language, therefore reducing the amount of similarity insights that reflect the use of language rather than similarity insights that are relevant to the user.

\begin{equation}
    Weight = f_1 \times \log\left(\frac{f_1}{f_e}\right) + f_2 \times \log\left(\frac{f_2}{f_e}\right)
\label{logarithmic_scaling}
\end{equation}

The weights can be refined with thresholding based on average frequency. That is, if the weight is smaller than the threshold, the weight is set to zero. The intuition of the rule is that if the word is more frequent in the English language than it is in any of these documents, that word might not be able to provide useful insights about these documents. The thresholding is required to eliminate common terms that were not meaningful in the distinction across documents.

\textbf{\begin{align}
    threshold &= \min(f_1, f_2, f_e) \\
    weight &= \frac{\log(f_1 + 1) \times e^{f_2}}{\sqrt{f_e} + 1} \\
    & if f_e > threshold
    :\, weight = 0 
\label{thresholding}    
\end{align}}

The final formula combines frequency differences, entity presence, and noun chunk recognition for an enriched analysis:

\textbf{\begin{equation}
    Weight = (\Delta f_1 + \Delta f_2) \times Base Multiplier
\end{equation}}

\noindent where:
\begin{itemize}
    \item Base Multiplier = \( 1 + 0.5 \times (\text{Entity Presence} + \text{Noun Chunk Presence}) \)

        \item \(\Delta f_1\) and \(\Delta f_2\) are the squared differences between the n-gram's frequency in documents 1 and 2 compared to the English corpus frequency:
        \begin{align*}
            \Delta f_1 &= (f_1 - f_e)^2 \times Normalization Factor \\
            \Delta f_2 &= (f_2 - f_e)^2 \times Normalization Factor
        \end{align*}
        \item Entity Presence is 0.5 if the n-gram is a named entity.
        \item Noun Chunk Presence is 0.5 if the n-gram is a noun chunk.
  
\end{itemize}

\begin{figure}[ht]
    \centering 
    \includegraphics[scale=0.5]{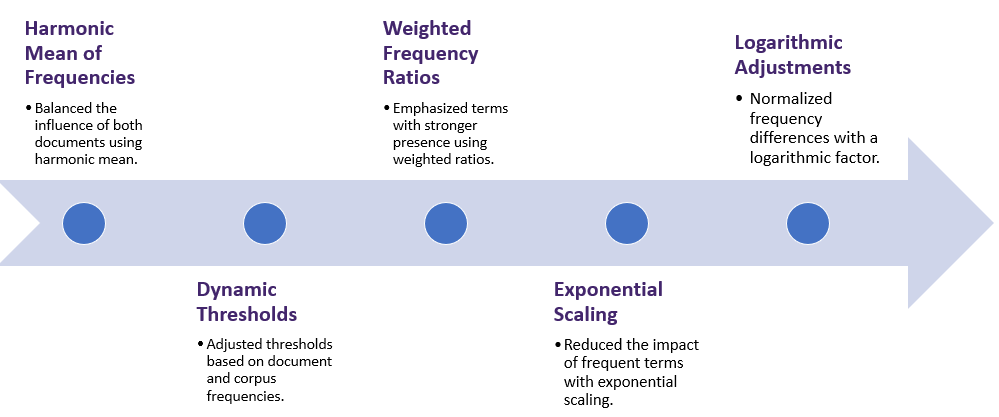} 
    \caption{Iteration of the analysis.}
    \label{fig:iterations}
\end{figure}

This formula synthesizes linguistic features and frequency adjustments to provide a more meaningful representation of textual similarity. If the n-gram contains an entity or noun chunk, the weight assigned to the n-gram will be increased. Rather than relying on raw counts alone, it makes the comparison meaningful beyond mere frequency matching, allowing to prioritize n-grams that are more likely to be meaningful to the user. Figure~\ref{fig:iterations} illustrates the steps of the analysis.

The intuition of the method is that it identifies topics that are common to the two text documents, but are not common in the English language. That means that the formula assigns higher weights to words that appear more frequently in both documents, but penalizes words that are frequent in other documents that are not related to the documents or topics that are being compared. The frequency of words in other documents is determined by their frequency in the English language.   

For instance, words such as `'is'' or ``the'' that are common between virtually any two English documents will not be assigned with high weights by the method since these words are also common in the English language in general. That leads to high weights assigned to topics that are relevant to the specific similarities between the two documents. Words that are less common in the English language but appear in both documents in relatively high frequency will be assigned higher weights, and therefore ore likely to be identified. 

Because the formula is deterministic, and the reason for the weight assigned to each word is clear, the model is explainable. That is different from methods based on deep neural networks such as large language models, where the rules are highly complex and non-intuitive, and users or developers of these methods cannot fully understand how the output is generated. The results are visualized as a word cloud, were the cloud is driven by the weight assigned to each word. That provides an intuitive and user-friendly visualization, as will be demonstrated in Section~\ref{results}.

\section{Experimental results}
\label{results}

The method was tested using a variety of text documents. While the focus was primarily on biographical texts and classical literature for testing, our methodology is not dependent on training or pre-existing knowledge, and therefore flexible enough to work with different types of content. This adaptability helps to handle different writing styles and different topics without the need to re-train the algorithm or adjust its parameters.

Main sources of the documents were Wikipedia articles and classical texts from Project Gutenberg. The biographies included the Wikipedia articles for Michael Jackson (9,603 words), Lionel Messi (21,465 words), Cristiano Ronaldo (12,784), Stephen Hawking (9,119 words), Taylor Swift (9,174 words), Jeff Bezos (7,279 words), and Albert Einstein (13,116 words). Classical text included ``War and Peace'' (534,562 words) and ``Anna Karenina'' (353,828 words).

% Donald Trump (17,874 words), 
% From the classical literature domain, Shakespeare's Hamlet (37,232 words) and Macbeth (20,626 words) were selected from Project Gutenberg \citep{stroube2003literary}. These texts offer a rich contrast to the modern language found in the Wikipedia entries, allowing us to see how well our approach works across different linguistic styles. In summary, the dataset used in this research spanned over different domains and linguistic styles to test the adaptability of the method as a non-parametric approach. 

The results of the method were compared to the results generated by a large language model implemented by ChatGPT 3.5 \citep{chatgpt} as provided by {\it OpenAI}. ChatGPT 3.5 is based on deep neural networks implementing large language models trained with very substantial amount of text documents, primarily retrieved from the internet. The proposed method was also compared to the results of the commonly used Doc2Vec algorithm \citep{le2014distributed}. That was done by applying word similarities to the two documents with vector size of 1000 words and window size of 50. The similarity was applied by identifying the 50 words most similar to the first entity, and comparing the output to the 50 words that are the most similar to the second entity. The two lists of words are then compared to identify words that appear in both lists. Words that are among the most similar words to both entities, as determined by Doc2Vec, can be assumed to be words that are common to both entities.

\subsection{Comparison of Albert Einstein and Stephen Hawking}

This experiment analyzes the linguistic patterns from the biographies of Stephen Hawking and Albert Einstein, sourced from their Wikipedia pages. Hawking's biography comprises 9,919 words, while Einstein's contains 13,116 words. This comparison aims to highlight the distinct contributions of each physicist and the effectiveness of the n-gram analysis in capturing similarities between their biographies and scientific legacies.

\begin{figure}[ht] 
    \centering 
    \includegraphics[scale=0.5]{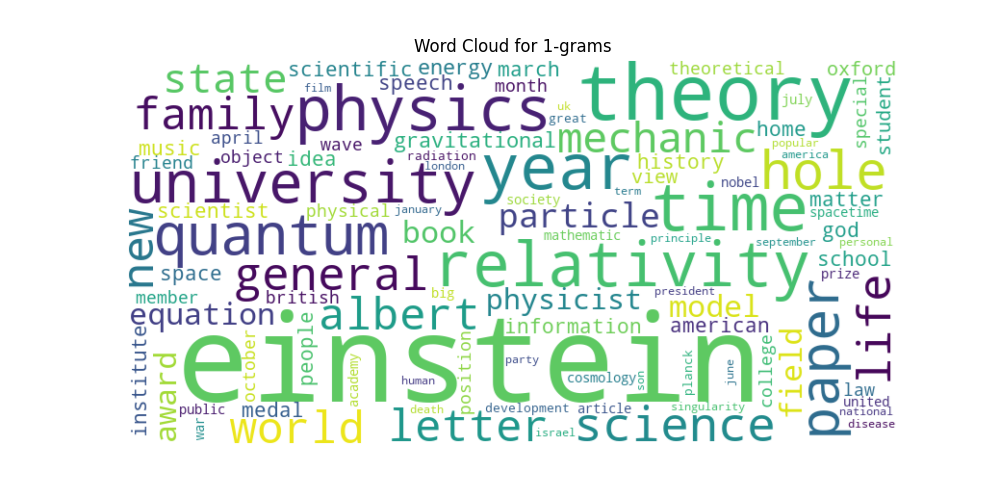} 
    \caption{Word cloud for the results of the Unigram  analysis of the similarities between for Stephen Hawking and Albert Einstein.} 
    \label{einstein_unigram} 
\end{figure}

Figure~\ref{einstein_unigram} shows the word cloud of the results of the unigram analysis applied to the two documents. The unigram analysis reveals key terms such as ``theory," ``universe," ``energy,"  and ``relativity," highlighting Einstein's work in relativity and Hawking's work based on the foundations of Einstein, and specifically on black holes. Terms such as ``paper", ``scientific'',  ``physics'', and ``university'' reflect that both persons were scientists, and specifically physicists. Terms such as ``energy" and ``time'' reveal their specific work related to physics. While these similarities between these two scientists are known, a person who is not aware of their background can make quick initial insights from the results. Words such as ``divorce'' and ``music'' are less trivial, but provide useful information since both men went through a divorce, and both were very interested in classical music. The method therefore helps to identify these similarities automatically, without the need to read and memorize facts mentioned in the documents, and then match facts that are common to both biographies.

The word cloud also shows some terms that do not provide useful insights such as ``state'' or ``family''. Such terms are provided by the equation, but do not provide any useful insight even for those who are not familiar with these scientists.

\begin{figure}[ht] 
    \centering 
    \includegraphics[scale=0.5]{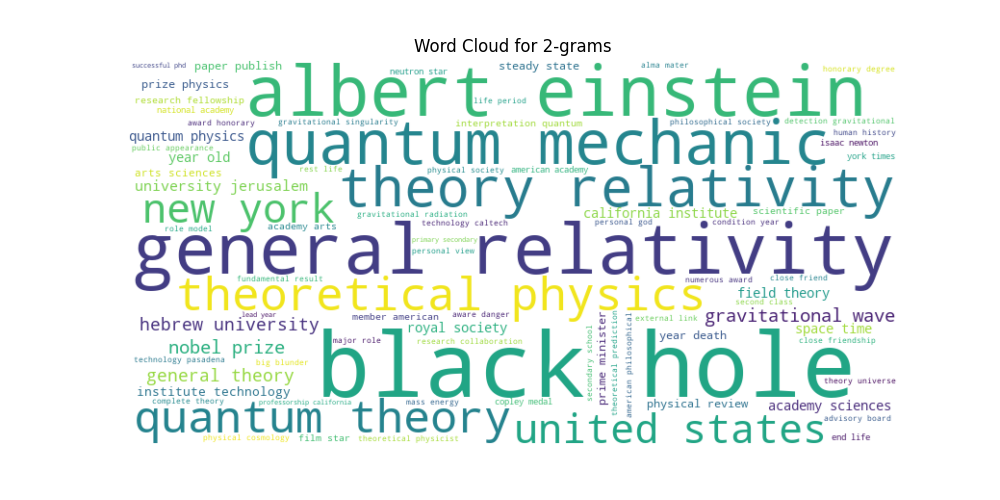} 
    \caption{Word cloud for the bigram analysis of the similarities between Stephen Hawking and Albert Einstein.} 
    \label{einstein_unigram} 
\end{figure}

Figure~\ref{einstein_unigram} shows the word cloud of applying the bigram analysis. The analysis provide insights such as ``general relativity", ``quantum mechanic'' and ``black hole", as these topics clearly were of interest to both scientists. ``Nobel prize'' is also mentioned although Hawking did not win the award, but was still mentioned as a candidate. Some academic societies where both scientists were members are also mentioned.

\begin{figure}[ht] 
    \centering 
    \includegraphics[scale=0.5]{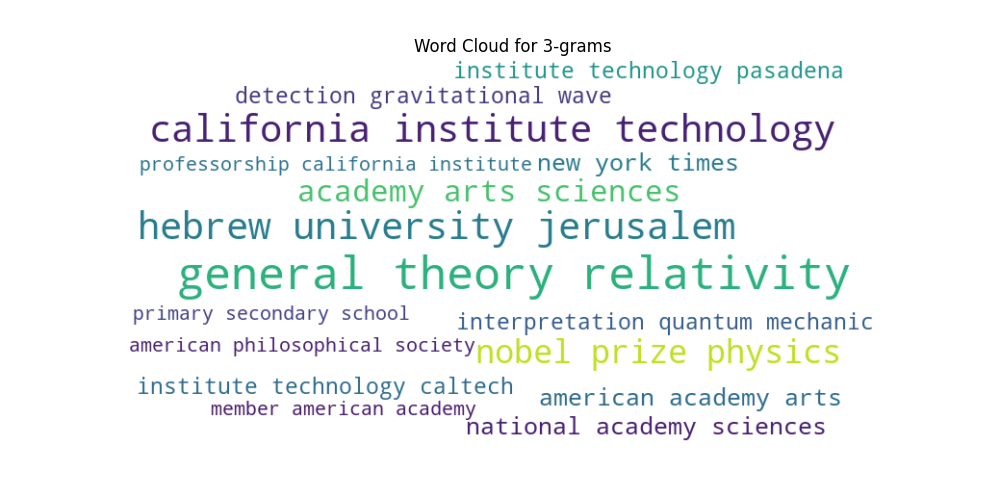} 
    \caption{Word cloud of the trigram analysis for Stephen Hawking and Albert Einstein.} 
    \label{einstein_trigram} 
\end{figure}

Figure~\ref{einstein_trigram} shows the resulting word cloud of the trigram analysis. As the figure shows, it highlights phrases such as ``general theory relativity", and several societies or institutions where the two spent time such as Caltech or the Hebrew University of Jerusalem. Einstein was a founding father of the Hebrew University of Jerusalem, while Hawking lectured there, including during the university's 70th anniversary in 2009. Both were elected to the American Philosophical Society—Einstein in 1937 and Hawking in 2008—for their significant scientific contributions. They were also affiliated with the California Institute of Technology, where Einstein predicted gravitational waves, and Hawking advanced the understanding of cosmic phenomena like black hole mergers. Caltech was also pivotal in the first detection of gravitational waves in 2015.

The results from Figures~\ref{einstein_unigram} through~\ref{einstein_trigram} can be compared to a request from ChatGPT 3.5 to provide the similarities between Stephen Hawking and Albert Einstein, following the question ``"How do Stephen Hawking and Albert Einstein compare in terms of their contributions to physics?". The summary made by ChatGPT large language model is the following:

\begin{mdframed}[linewidth=0.5pt, roundcorner=5pt] 
    \begin{quote} 
    \textit{"Stephen Hawking and Albert Einstein are famous physicists. Einstein is known for his theory of relativity, and Hawking is known for his work on black holes. They both made important contributions to physics."} 
    \end{quote} 
\end{mdframed}

While this summary provides a basic overview that is correct, it mostly summarizes the work of each scientist. ChatGPT therefore can focus on the major similarities, but might struggle to identify specific details of the similarities that can be deduced by very careful reading of their brief biographies. The summary provided by the large language model therefore provides just little directions for a user interested in exploring the similarities between the two persons.

When using the most similar words as determined by Doc2Vec, the words among the top 50 similar words that appear for both Einstein and Hawking are ``He'', ``his'', ``the'', ``and'', ``from'', ``by'', ``''it'', ``new'', ``that'', ``first'', ``physics'', ``would'', ``including'', ``on'', ``jewish'',  ``were'', ``is'', ``with''. That list mostly includes common English words that do not necessarily provide useful information. Since Doc2Vec does not penalize words that are frequent in the English language, the high frequency of these words can lead Doc2Vec to associate them with the entities, therefore leading the algorithm to select these words. The word ``physics'' provides the information that both were physicists. The word ``new'' might indicate that both were innovative. The word ``jewish'' also appears as a common word for both people, although Stephen Hawing was not Jewish.  The male pronouns can indicate that both entities were people, and both were men.

\subsection{A comparison between soccer players Cristiano Ronaldo and Lionel Messi}

\begin{figure}[ht] 
\centering 
\includegraphics[scale=0.65]{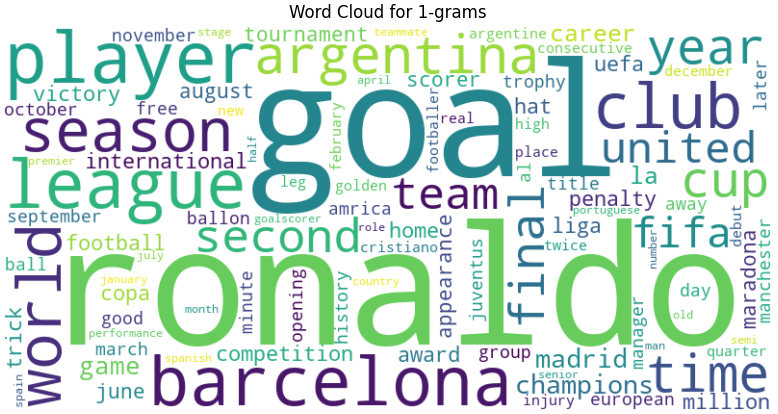} 
\caption{Word cloud of the results of the unigram comparison between Cristiano Ronaldo and Lionel Messi.}
 \label{players_unigram}
\end{figure}

This study analyzes the patterns of similarities from the biographies of Cristiano Ronaldo and Lionel Messi, as taken from their Wikipedia articles. Figure~\ref{players_unigram} shows the word cloud that results from the comparison between Messi and Ronaldo as reflected by the 1-gram analysis. As in the experiment with Einstein and Hawking, the insights gained from n-gram analysis are also compared with those generated by the large language model provided by ChatGPT. 

Figure~\ref{players_unigram} shows the word cloud of the results of the unigram analysis of the similarities between Ronaldo and Messi. The analysis provides terms such as ``goal," ``team," ``league," ``football'', ``ball'', and ``match," showing the user that the two are soccer players. The analysis shows the tournaments the two played in, countries, teams, and awards they received. For instance, the term ``Ballon'' reveals that the two hold the record in winning and nomination for Ballon d'Ors. Naturally, soccer experts might not be surprised from these results, but it demonstrates that the method can identify similarities through terms that lead to insights about the players. Such terms can be used by the user as insight to get further information about possible similarities they were not aware of initially. 

\begin{figure}[ht] 
\centering 
\includegraphics[scale=0.65]{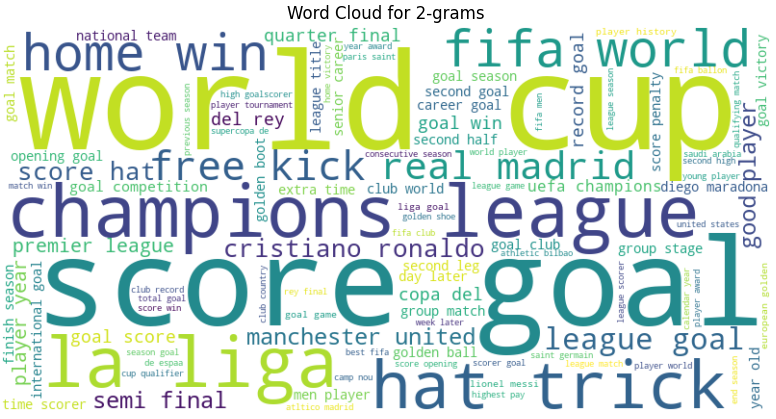} 
\caption{Word cloud of the results of the bigram comparison between Cristiano Ronaldo and Lionel Messi.} 
\label{players_bigram}
 \end{figure}

Figure~\ref{players_bigram} shows the world cloud of the bigram analysis, revealing terms related to soccer, including ``Champions League” and ``la liga'', two leagues were both soccer players played. Terms like ``total goals'' and ``club record'' indicate that both players hold the scoring record for their respective soccer clubs. ``World cup'', ``fifa world'', and ``del rey'' indicate on some of the tournaments both players competed in. ``Diego Maradona'' did not play at the same time as these players, but both were often compared to him. ``Saudi Arabia'' is mentioned as both players received offers to play in that country, and Ronaldo did play in the Saudi Arabia soccer league. ``Golden Boot'' is another term that the two players share, as both wan that award. Terms such as ``Manchester United”, ``Real Madrid”, and ``FC Barcelona” highlight the teams where the two players played against each other. 

\begin{figure}[ht] 
\centering 
\includegraphics[scale=0.65]{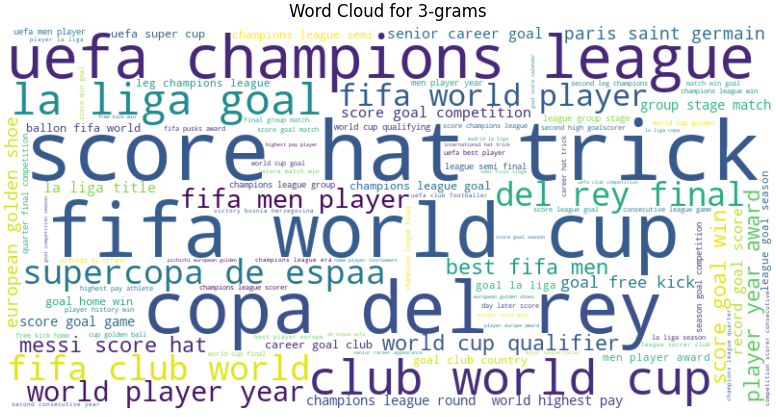} 
\caption{Cloud word of the trigram analysis of the similarity between Cristiano Ronaldo and Lionel Messi.} 
\label{players_trigram} 
\end{figure}

The trigram analysis of the two soccer players is shown by Figure~\ref{players_trigram}. Phrases like ``Golden Boot winner” and ``World Cup final” highlight achievements that are common to both players. ``copa del rey'', ``uefa champions leagues'', and ``fifa world cup'' mostly refer to competitions where both players played. Terms such as ``highest paid player'' refer to the fact that the two players were the highest paid players during their career. For a person not familiar with these two figures, that information can provide useful insights that can be used as a base for a more thorough investigation.

The large language model by ChatGPT being asked to provide a comparison between the two players by the question ``How do Lionel Messi and Cristiano Ronaldo compare in terms of their football careers?". It provided the following basic summary:\\

\begin{mdframed}[linewidth=0.5pt, roundcorner=5pt] \begin{quote} \textit{"Cristiano Ronaldo and Lionel Messi are two of the greatest football players of all time. Ronaldo is known for his athleticism, while Messi is known for his playmaking and dribbling. They have won numerous awards and broken many records in football."} \end{quote} \end{mdframed}

While this summary provides a basic overview, and beyond a quick summary of each of the two players. Yet, it lacks specific terminology and nuances critical for a comprehensive understanding of the specific achievements and career stages that are common to the two players. The large language model therefore provides a correct overview, but that overview does not provide insights that can be used for further and more detailed understanding of the similarities between the two subjects.

Analysis of the 50 most similar words to both players provided several words that were in both Messi's and Ronaldo's lists. These included words such as ``a'', ``on'', ``at'', ``the'', ``number'', ``player'', ``him'', ``in'', ``became'', ``for'', ``by'', ``shirt'', ``years'', ``their'', ``with'', ``Ronaldo's'', ``new''. The list of words that are common in both lists of most similar words shows some indication that the two are athletes. But since it is saturated by words that are frequent in the English language and have no informative meaning, they do not provide many useful insights about the similarities between the two players. 

\subsection{Comparison between pop icons Michael Jackson and Taylor Swift}

\begin{figure}[ht] 
\centering 
\includegraphics[scale=0.65]{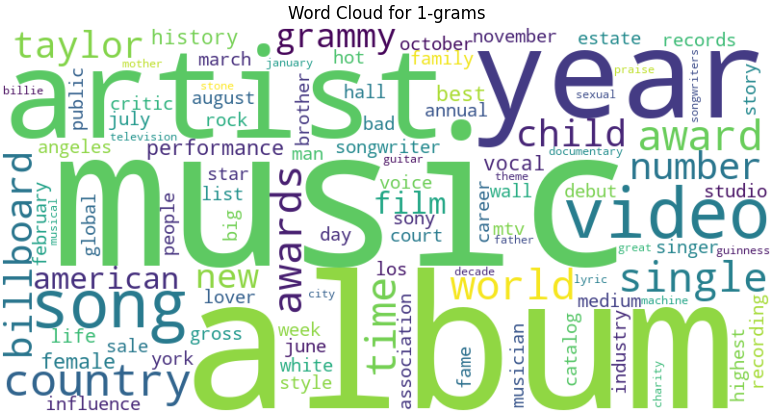} 
\caption{Cloud word of the unigram analysis of the similarity between Michael Jackson and Taylor Swift.} 
\label{jackson_swift_ngram_1_wordcloud}
 \end{figure}

Figure~\ref{jackson_swift_ngram_1_wordcloud} shows the word cloud resulting from comparing the Wikipedia articles describing Michael Jackson and Taylor Swift. Terms such as ``music, ``singer'', ``album," ``song," ``video'', ``artist'', ``performance'', ``records'', ``voice'', ``studio'', ``mtv'',  and ``music," indicate that the two are musicians and singers. Words such as``songwriter'' identify another similarity, as the two normally wrote their own songs. The word  ``court'' indicates that both musicians tended to be involved in legal affairs. The word ``Sony'' reveals that both musicians made major deals with Sony Music Group, and both signed with Sony. The word ``American'' indicates the similarity that both singers share the same nationality. ``Grammy'' indicates that both musicians won the Grammy award, which is another commonality between them. The word ``bad'' seems incidental, as both musicians had highly popular songs with the word ``bad'' in them.

\begin{figure}[ht] 
\centering \includegraphics[scale=0.65]{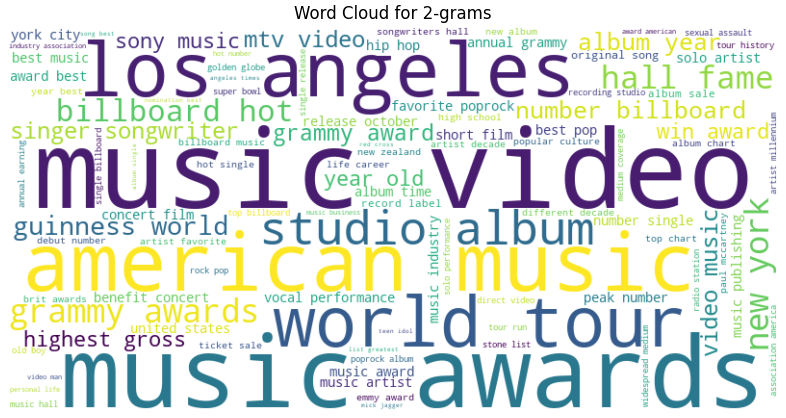} 
\caption{Cloud word of the bigram analysis of the similarity between Michael Jackson and Taylor Swift.} 
\label{jackson_swift_ngram_2_wordcloud}
\end{figure}

Figure~\ref{jackson_swift_ngram_2_wordcloud} shows the resulting word cloud of the bigram analysis. ``Grammy Award” and ``Billboard Chart” are naturally common to the successful musicians, as both won the Grammy Award and both had songs in the Billboard Chart. The term ``Los Angeles'' reveals that the two musicians lived in the same city. The term ``Guinness World'' is another interesting insight, as both musicians hold numerous Guinness records. For instance, Jackson is the record holder for being the best-selling artist of all time, while Swift holds the record for highest-grossing music tour among others.

\begin{figure}[ht]
 \centering \includegraphics[scale=0.65]{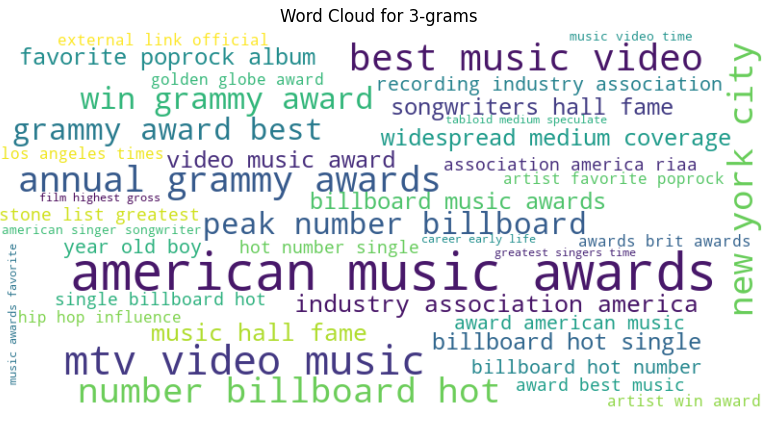} 
 \caption{Cloud word of the trigram analysis of the similarity between Michael Jackson and Taylor Swift.}
  \label{jackson_swift_ngram_3_wordcloud}
  \end{figure}

The trigram analysis is displayed by Figure~\ref{jackson_swift_ngram_3_wordcloud}. The diagram shows awards that both musicians won, and are therefore common to both musicians. These awards include the best video award, the Grammy award, Golden Globe, and the songwriters hall of fame. These terms show that the analysis can identify similarities between documents in a manner that is non-parametric yet clear.

When asked to compare the careers of Michael Jackson and Taylor Swift, ChatGPT generated the following basic summary:\\

% "How do Michael Jackson and Taylor Swift compare in terms of their impact on the music industry?"

\begin{mdframed}[linewidth=0.5pt, roundcorner=5pt] \begin{quote} \textit{"Michael Jackson was a legendary performer known as the King of Pop, and Taylor Swift is a singer-songwriter known for her storytelling. Jackson is famous for his iconic albums and dance moves, while Swift is recognized for her ability to connect with fans through her lyrics."} \end{quote} \end{mdframed}

While this summary provides a basic overview, but it lacks direct identification of similarities between them, and in fact highlights differences between the two musicians. A person who is not familiar with any of these musicians will therefore struggle to start a comprehensive analysis of the similarities based on this information alone.

Analysis using Doc2Vec to identify common terms that are similar to both artists led to the following words: ``has'', ``year'', ``a'', ``2024'', ``time'', ``the''. This is a basic list that does not provide information that allows to learn about the similarities between the artists. When applied to each artists, Doc2Vec provides  relevant terms. For instance, the list of more similar words to Taylor Swift included words like ``album'', ``Nashville'', ``woman'', ``country'', ``songs'', ``Billboard'', ``songwriter'', ``musician'', ``singer'', and ``pop''. But Doc2Vec did not find any of these words among the top words similar to Michael Jackson in his Wikipedia article, and therefore these words were not highlighted as insights of similarities between the two artists.

\subsection{Comparison of Albert Einstein and Jeff Bezos}

\begin{figure}[ht] 
\centering \includegraphics[scale=0.6]{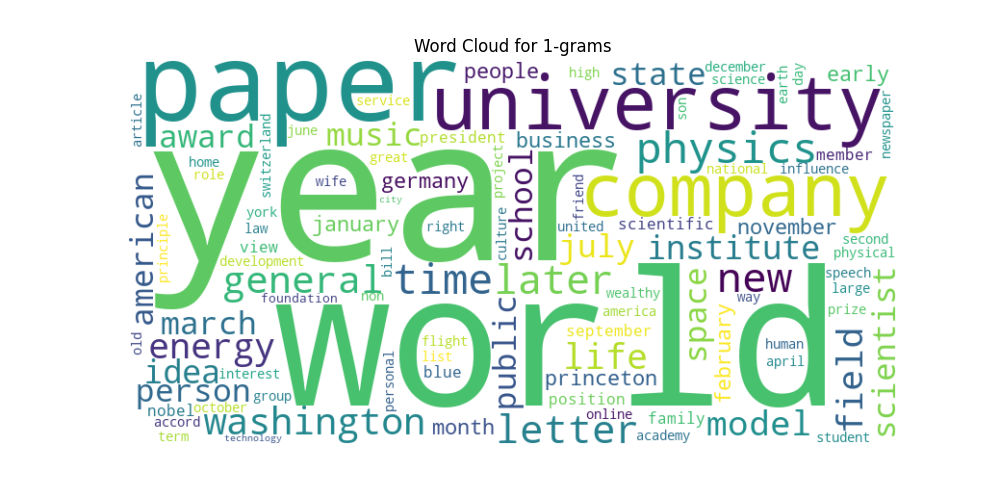} 
\caption{Word cloud generated by the unigram analysis of the comparison between Albert Einstein and Jeff Bezos.} 
\label{bezos_einstein_unigram} 
\end{figure}

The comparisons above were made between topics that share known commonalities. To test a situation where the topics do not share obvious commonalities, we compared Jeff Bezos to Albert Einstein. A business person such as Jeff Bezos is not expected to have obvious things in common with a science figure such as Albert Einstein. Figure~\ref{bezos_einstein_unigram} shows the unigram analysis of the similarities of Einstein and Bezos. 

The analysis provides similarities that might not necessarily be well-known. For instance, the term ``Princeton'' indicates that Jeff Bezos earned a Bachelor's degree from Princeton University, and Einstein taught at that institution. The word ``Germany'' could provide an insight that Einstein lived part of his early life in Germany, while Bezos have substantial business connections with Germany and he is even a member of the Germany Logistics Hall of Fame. The word ``space'' can provide insights that Einstein studied space, while Bezos visited it and has substantial business interest in it. The word ``American'' identifies their common nationality. The word music indicates that Einstein was highly interested in classical music, while there are several songs about Jeff Bezos, perhaps most notably by musician Bo Burnham.

\begin{figure}[ht] 
\centering \includegraphics[scale=0.6]{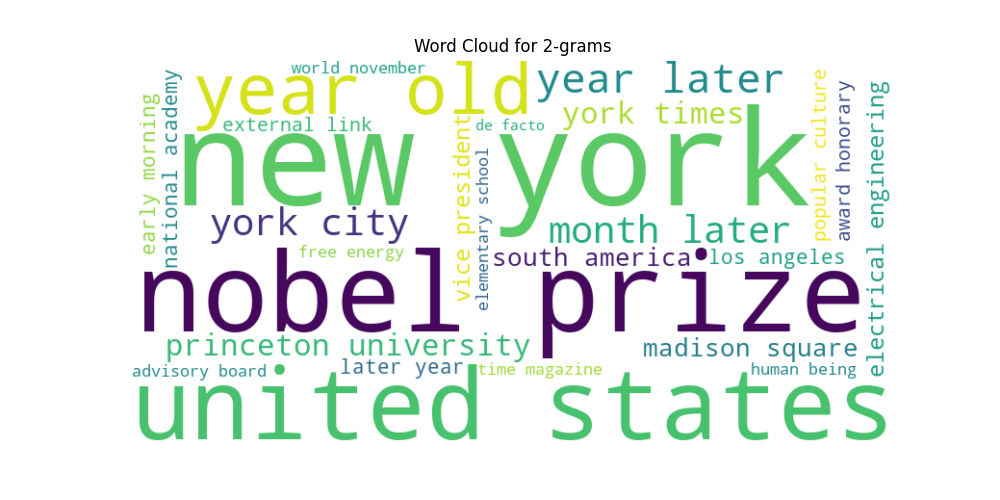} 
\caption{Word cloud generated by the bigram analysis of the comparison between Albert Einstein and Jeff Bezos.}
 \label{bezos_einstein_bigram} 
 \end{figure}

The bigram analysis displayed by Figure~\ref{bezos_einstein_bigram} shows terms that identify their nationality and places where the two lived. These include ```United States'', ``Los Angeles'',  and ``New York''. The topic ``electrical engineering'' indicates that Einstein had revolutionary impact on that field, while Bezos studied in the department of electrical engineering at Princeton University. Einstein was awarded a Nobel Prize, while Bezos funded Nobel Prize winner Shinya Yamanaka to study longevity and aging. The topic ``National Academy'' can also be considered surprising, but provides insights for similarities. Einstein was a member of the National Academy of Science, and Bezos is a member of the National Academy of Engineering.

ChatGPT was used to identify the similarities between Albert Einstein and Jeff Bezos through the question ``"How do Albert Einstein and Jeff Bezos compare in terms of their careers and achievements?"'. ChatGPT generated the following basic summary:\\\\

\begin{mdframed}[linewidth=0.5pt, roundcorner=5pt] \begin{quote} \textit{"Albert Einstein was a scientist known for his theories, and Jeff Bezos is the founder of Amazon. Einstein won awards for his work in physics, and Bezos is one of the richest people in the world because of Amazon."} \end{quote} \end{mdframed}

While this summary provides a basic overview, it basically provides a brief summary of each person. It does not identify the similarities between them. \\

\subsection{Comparison between ``War and peace'' and ``Anna Karenina'' }

The proposed method can be used for the field of Digital Humanities. Digital; humanities enables to studying of the humanities with the aid of digital and computational tools. The method shown here can identify not merely informative insights, but also words and phrases that are unique to two different creations. That can reveal insights regarding the selection of words and phrases that make the style of books or authors.

\begin{figure}[ht] 
\centering \includegraphics[scale=0.65]{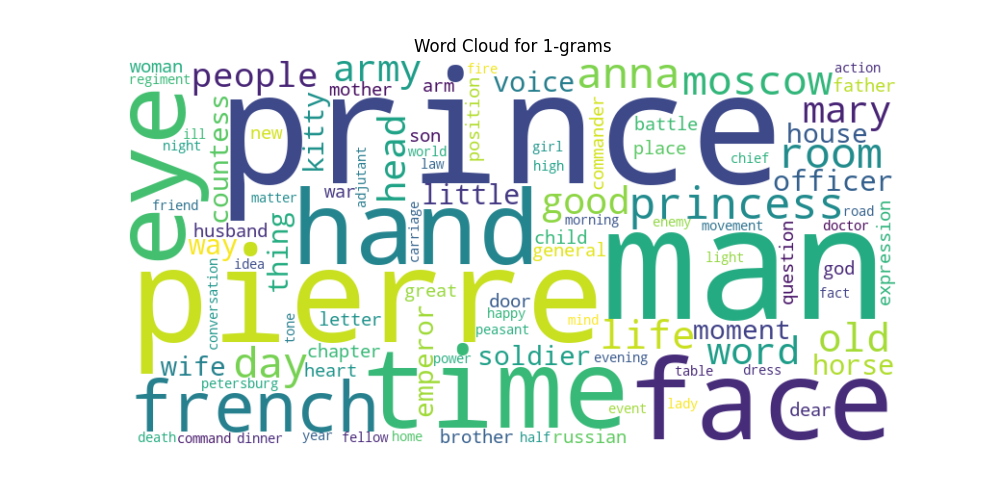} 
\caption{Word cloud generated by the unigram analysis of the comparison between Tolstoy's ``Anna Karenina'' and ``War and Peace''.}
 \label{tolstoy_1_gram} 
 \end{figure}

Here we tested two of Tolstoy's masterpieces: ``War and peace'' and ``Anna Karenina''. Figure~\ref{tolstoy_1_gram} shows the word cloud of the unigram analysis. The word cloud shows words such as ``prince'', that could be explained by the characters in the two books -- Prince Andrei Nikolayevich Bolkonsky in ``War and Peach'' and Prince Stepan Arkadyevich Oblonsky in ``Anna Karenina''.  ``French'' is expected because of the war against France in ``War and Peace'', and Anna Karenina often spoke French, also to display her social status. 

Some names were also common. The name ``Anna'' is mentioned in ``War and Peace'' as socialite Anna Pavlovna Scherer and Princess Anna Mikhailovna Drubetskaya. The name ``Pierre'' is Pierre Bezukhov in ``War and Peace'', and Petritsky in ``Anna Karenina'', who was addressed as ``Pierre''. But while the common names are expected, the common words and phrases show similarities in style and concepts emphasized in the books, and might be of interest to humanists. 

\begin{figure}[ht] 
\centering \includegraphics[scale=0.65]{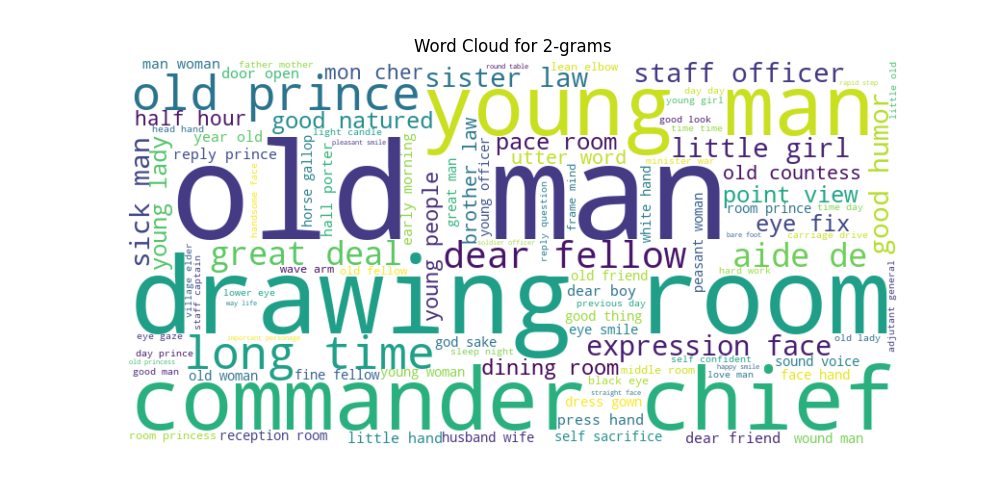} 
\caption{Word cloud generated by the bigram analysis of the comparison between Tolstoy's ``Anna Karenina'' and ``War and Peace''.}
 \label{tolstoy_2_gram} 
\end{figure}

For instance, the 2-gram word cloud shown in Figure~\ref{tolstoy_2_gram} highlights  terms related to age such as ``old man'', ``young man'', ``little girl'', ``young girl'', ``young lady'', ``year old'', ``young officer'', ``old countess'', ``old prince'', ``young people'', and ``old woman'', showing focus on age in Tolstoy's work. Similarly, the world cloud shows a large number of terms related to family relations or parts of a house, which might be of interest to those trying to study and profile Tolstoy's style. 

\begin{figure}[ht] 
\centering \includegraphics[scale=0.65]{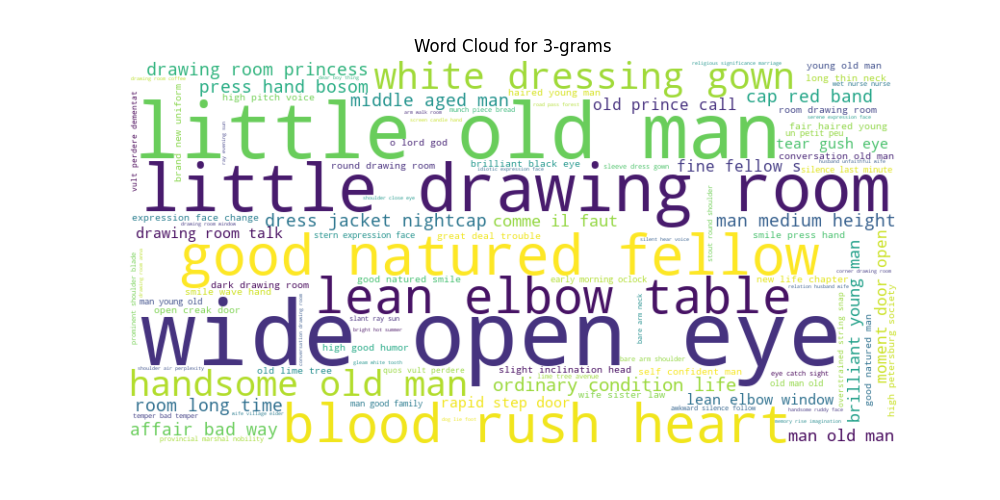} 
\caption{Word cloud generated by the trigram analysis of the comparison between Tolstoy's ``Anna Karenina'' and ``War and Peace''.}
 \label{tolstoy_3_gram} 
\end{figure}

Figures~\ref{tolstoy_3_gram} shows the 3-gram analysis, showing combinations used by Tolstoy in both books. While they do not necessarily provide information about the two books, they show the wording common to Tolstoy's two books, and can reveal information about Tolstoy's unique style and wording. These patterns are difficult to notice by manual reading of the books since most combination of words are common and are not unique to a certain author.

\section{Conclusion}
\label{conclusion}

The ability to provide an analysis of similarities between  two topics or subjects of interest can assist in knowledge extraction. Reading and comparing documents is a time-consuming task, making it more difficult for humans to perform such comparisons without automated assistance. When the documents are large, such comparison is also limited by the capacity of the memory of the person trying to perform the task.

Here we describe a method that can perform the task in an explainable and non-parametric manner. It reads directly from the documents without the need to train a machine learning system, and provides a list of recommended insights of possible similarities. While the list may contain terms that are not necessarily similarities between the subjects, it also provides terms that lead to similarities that are difficult to identify manually. The method is general, and can be applied to any kind of text. 

The solution is based on a defined mathematical formula that assigns higher weight to terms that appear in the two document, but penalizes terms or words that are common in the English language regardless of the topic. The output of the method is therefore explainable, and the user can understand the reasons for selecting each of these terms. That makes it different from solutions that are based on large language models, where the rules are data-driven and highly complex, and a user cannot practically understand why a certain answer is given. It is also different in that sense from Doc2Vec, which also uses a neural network in its analysis. The proposed method is certainly different from document comparison methods that are commonly used by word processing applications. These document comparison methods aim at identifying text differences between the documents rather than topics or patterns of similarity.

The output of the method is not necessarily a readable paragraph, but a set of weighted terms visualized through a word cloud. Each term can be considered a possible insight, and the user can therefore observe these terms and explore them further to learn to full nature of these similarities. In that sense the proposed method is used to assist handling large amounts of text document. 

While large language models can also be used to analyze similarities between documents, the output provided by these models is different from the results of the analysis shown here. Large language models can provide readable paragraphs, but at the same time it is more difficult to understand how these models are producing their output. It is therefore more difficult to understand how complete or comprehensive the analysis is.  

The proposed method is limited to common terms, and therefore cannot identify more complex patterns that require deep understanding of the subjects being studied. It therefore cannot fully mimic the way humans identify similarities between subjects. The goal of the proposed method is to provide a tool that can aid a person studying the similarities between two topics. In that sense it is a solution that can be used in combination with other methods to provide quick and explainable insights to the user. In its present form the solution is limited to the English language, although adjustments to other languages can be implemented by using similar NLP tools designed for each target language. Another limitation of the method is that its performance depends on NLP tools that perform tasks such as tokenization and lemmatization. These tools are not necessarily perfect, and their imperfectness might affect the performance of the propose method.

\section*{Data availability}

The code that implements the method is available at \url{https://github.com/AkhilJoshi15/Language_Pattern_Analyzer}.

\section*{Conflict of interest statement}

The authors declare no conflict of interest.

\section*{Acknowledgments}

This study was supported in part by NSF grant 2148878. We would like the thank the three knowledgeable reviewers for the insightful comments.

\bibliography{main}

\end{document}